\documentclass[aps,pra,superscriptaddress,twocolumn]{revtex4}
\usepackage[T1]{fontenc}
\usepackage[utf8]{inputenc}
\usepackage{amsfonts}
\usepackage{amsmath}
\usepackage{graphicx}
\usepackage{color}
\usepackage{amssymb}
\usepackage[normalem]{ulem}
\usepackage{siunitx}
\usepackage{upgreek}
\usepackage{braket}
\usepackage{bm}
\DeclareMathAlphabet{\mathbbold}{U}{bbold}{m}{n}
\newcommand{\ve}{\varepsilon}
\newcommand{\e}{\text{e}}

\renewcommand{\i}{\text{i}}
\renewcommand{\d}{\text{d}}

\begin{document}

\title{Rotational Quantum Friction via Spontaneous Decay}

\author{Nicolas Sch\"uler}
\email{nicolas.schueler@physik.uni-kassel.de}
\affiliation{Institut f\"ur Physik, Universit\"at Kassel, Heinrich-Plett-Stra\ss e 40, 34132 Kassel, Germany}

\author{O. J. Franca}
\email{uk081688@uni-kassel.de}
\affiliation{Institut f\"ur Physik, Universit\"at Kassel, Heinrich-Plett-Stra\ss e 40, 34132 Kassel, Germany}

\author{Michael Vaz}
\affiliation{SPEC, CEA, CNRS, Université Paris-Saclay, 91191, Gif-sur-Yvette, France}

\author{Herv\'e Bercegol}
\affiliation{SPEC, CEA, CNRS, Université Paris-Saclay, 91191, Gif-sur-Yvette, France}

\author{Stefan Yoshi Buhmann}
\email{stefan.buhmann@uni-kassel.de}
\affiliation{Institut f\"ur Physik, Universit\"at Kassel, Heinrich-Plett-Stra\ss e 40, 34132 Kassel, Germany}

\begin{abstract}
\noindent
A fascinating effect belonging to the field of vacuum forces and fluctuations is that of quantum friction. It refers to the prediction of a dissipative force acting on a moving object due to the quantum vacuum field. In this work, we investigate rotational quantum friction where a diatomic polar molecule rotates around its own center of mass in free space. We quantize the rotational motion and investigate the resulting dissipation due to spontaneous decay. We find in the Markovian regime that a friction torque $\propto \Omega^3$ persists even for zero temperature, and in agreement with the classical result in the limit of large rotational quantum number $l$. Within the non-Markovian short-time regime we find a friction $\propto\Omega$.

\end{abstract}

\maketitle

\textit{Introduction---}Since quantum friction was first correctly proposed in 1997 \cite{Pendry1997}, it has witnessed a great deal of interest, both in favor of and against the very idea. Within a period of roughly 25 years, the research field of quantum friction has branched into different friction types with many different setups, even though it has not been observed experimentally to date. Casimir-type quantum friction, which involves two macroscopic plates, has been named after the Casimir effect \cite{Casimir1948a}.
Initially causing quite a heated debate
\cite{ Philbin2009, Pendry2010a, Leonhardt2010, Milton2016, Pendry2010b}, Casimir-type quantum friction has been investigated in multiple scenarios \cite{Volokitin1999, Volokitin2003, Volokitin2008, Volokitin2011, Mkrtchian1995}.

Casimir--Polder-type quantum friction is the dissipative force between a macroscopic body and a moving atom, and is named after the work by Casimir and Polder \cite{Casimir1948b}, who calculated the force that a perfectly conducting plate exerts on a static atom. Despite its relatively simple setup, different results have been obtained regarding the velocity dependence of the friction force \cite{Scheel2009, Intravaia2014}, ranging from $v$ to $v^3$. The discrepancy can be traced back to the use of different approximations, namely linear response theory versus the Markov approximation \cite{Klatt2021}. A complementary approach was presented in Ref.~\cite{Sinha}, where the interaction between a polarizable particle and blackbody radiation field, and the implications for energy conservation were analyzed, thereby linking the phenomenon to diffusion.

Owing to the limited observability of the Casimir--Polder-type quantum friction, we turn to the experimentally more accessible case of rotational quantum friction \cite{Manjavacas2010}. It can be investigated in stationary setups involving either (i) rotating macroscopic bodies, such as a disc \cite{Decca}, or (ii) rotating microscopic objects, including pair of neutral atoms \cite{Bercegol,Vaz}  or nanospheres \cite{Manjavacas2012,Manjavacas2017}. The latter configuration can be experimentally realized using optical centrifuges \cite{Karczmarek}, which provide powerful control over molecular rotation and enable the study of molecular dynamics and properties at extreme levels of rotational excitation \cite{Milner2014,Milner2019,Milner2020}.

In our work, we consider a rigid rotor consisting of two ions, from here referred as molecule. We assume it to be neutral but polar and to rotate around its center of mass with angular velocity $\bm{\Omega}$. We will analyze the interaction of the molecule with the quantum vacuum by investigating the dynamics of $\bm{\Omega}(t)$. Our approach differs from previous work by quantizing the angular velocity via the orbital angular momentum quantum number $l$. Since the rotating molecule represents a fluctuating dipole which is subjected to photonic decay by spontaneous emission, its velocity will eventually decrease as a result of the energy loss. This decrease in $\bm{\Omega}$ is quantized in terms of $l$ and may be regarded as a friction but also as a normal spontaneous emission.

Our work is structured as follows: after formulating the rotational friction problem, we will first introduce a classical solution based on radiation reaction. We will then solve the quantum dynamics in terms of transition rates and connect our result to the classical one in the limit of large angular momentum.  

\textit{Dissipative torque---}In classical mechanics, the rotational dynamics of a rigid body is described by the angular momentum vector $\bm{L}$ and the angular velocity $\bm{\Omega}$ related by a linear transformation in the body frame \cite{Goldstein2002}
\begin{equation}\label{LIOmega}
    \bm{L}=\mathbb{I}\cdot\bm{\Omega}\;,
\end{equation}
where $\mathbb{I}$ is the moment of inertia tensor. Along a given principal axis of the body, the equations of motion are
\begin{equation}\label{Torque}
    N=\dot{L}=I\dot{\Omega}\;,
\end{equation}
where $N$ is the torque. The kinetic energy of motion about the axis of rotation can be expressed in terms of these quantities by means of Eq.~(\ref{LIOmega}) as
\begin{equation}
    E=\frac{1}{2}I\Omega^2=\frac{L^2}{2I}\;.
\end{equation}
Consequently, the instantaneous power lost by the rigid body reads \cite{Schaum}
\begin{equation}\label{RotPower}
    P=-\dot{E}=-N\Omega=-I\Omega\dot{\Omega}\;,
\end{equation}
where Eq.~(\ref{Torque}) has been used. By analogy with sliding friction, which is frequently linear in the velocity of the corresponding particle as given by Stoke's law \cite{Goldstein2002,Bird2002} or quadratic as known from the drag force for high-speed flows \cite{Taylor2005}, we assume a power-law friction for the rotating body
\begin{equation}
    N=-\alpha\Omega^k\;.
\end{equation}
This enables us to rewrite Eq.~(\ref{RotPower}) yielding
\begin{equation}
    P=-N\Omega=\alpha\Omega^{k+1}\;,
\end{equation}
where $\alpha$ denotes the friction coefficient and $k$ is a positive integer.

\begin{figure}[t]
    \centering
    \includegraphics[scale=0.8]{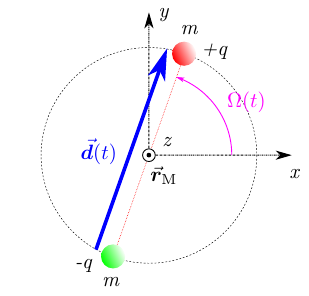}
    \caption{Two charges $+q$ and $-q$ of identical mass $m$ form a rigid, classical electric dipole $\bm{d}$ rotating in the electromagnetic vacuum around the $z$ axis at angular velocity $\bm{\Omega}$.}
    \label{Dipole and geometry}
\end{figure}

\textit{Classical description---}Let us consider a rigid, classical electric dipole rotating in the electromagnetic vacuum around the $z$ axis at angular velocity $\Omega$. We use the notation $\mu$ for the reduced mass of the dipole and $d$ for the constant norm of the rigid dipole moment (See Fig.~\ref{Dipole and geometry}). 
We assume that the two poles have a respective electric charge of $\pm q$. The dipole moment can then be written as:
\begin{equation}\label{Dipole}
    \bm{d}(t) = d \left[ \cos\left(\Omega t\right) \bm{e}_x + \sin\left(\Omega t\right) \bm{e}_y \right]\;,
\end{equation}
where $\bm{e}_x$ and $\bm{e}_y$ are two fixed and orthogonal unit vectors, the axis of the angular momentum is $\bm{e}_z = \bm{e}_x \times \bm{e}_y$ and the dipole's moment of inertia is given by $I=\mu d^2/q^2$. 

The classical dynamics of such rotating classical dipole is given by 
\begin{equation}
    \mu \frac{d^2}{q^2} \dot{\Omega} \bm{e}_z = \bm{d} \times \bm{E}_\text{rr}
\end{equation}
with the radiation reaction field
\begin{equation}
    \bm{E}_\text{rr} = \frac{\mu \tau}{q^2} \dddot{\bm{d}}
\end{equation}
characterized by the radiation reaction time:
\begin{equation}
    \tau = \frac{q^2}{6\pi \ve_0 \mu c^3}.
\end{equation}
Note that the electromagnetic vacuum field does not explicitly appear in 
this classical picture. %
The equation for the rotation speed is then given by
\begin{equation}\label{classical_damping_equation}
    \dot{\Omega} = - \tau \Omega^3 \;.
\end{equation}
This equation admits the solution \cite{Duviryak}
\begin{equation}
    \Omega(t) = \frac{\Omega_0}{\sqrt{1+2(\Omega_0 \tau)(\Omega_0 t)}} \;,
\end{equation}
which gives the damping on the rotation speed due to purely classical radiation. In particular, by computing the radiated power starting from the dynamical equation (\ref{classical_damping_equation}), we obtain
\begin{equation}
    P_{\mathrm{cl}} = - I \dot{\Omega} \Omega = \mu\frac{d^2}{q^2} \tau \Omega^4 = \frac{d^2}{6 \pi \ve_0 c^3} \left( \frac{L}{I} \right)^4\;,\label{Eq classical power}
\end{equation}
where Eq.~(\ref{LIOmega}) has been used to express this result in terms of the classical angular momentum $L$ of the dipole. As a consistency check this result can also be derived by means of Larmor's formula \cite{Larmor}
\begin{equation}
    P_\mathrm{Larmor} = \frac{2}{3} \frac{e^2}{4\pi\ve_0c^3} |\dot{\bm{v}}|^2 \;,
\end{equation}
where $\dot{\bm{v}}=-\frac{\Omega^2}{e}\bm{d}(t)$ stands for the acceleration of the nonrelativistic  charge.

\textit{Quantum description---}The rotational motion of the dipole of Fig.~\ref{Dipole and geometry}, now characterized as a diatomic polar molecule, is quantized via the rotational states $\ket{l}$, which are eigenstates of the angular momentum operator $\hat{\bm{L}}$:
\begin{equation}\label{L2}
    \hat{\bm{L}}^2\ket{l}=\hbar^2l(l+1)\ket{l}.    
\end{equation}
We describe the rotational motion via the average angular velocity $\Omega(t)$, which is a sum of the state-dependent velocities
\begin{equation}\label{Quantized Omega}
    \Omega_l=\frac{\hbar}{I}\sqrt{l(l+1)} \;,
\end{equation}
which depend on the occupied rotational quantum numbers $l$:
\begin{equation}\label{RotationalVelocity}
    \Omega(t) = \langle\hat{\Omega}(t)\rangle =\sum_lp_l(t)\Omega_l\;.
\end{equation}
Thus,
\begin{equation}\label{OmegaDot}    
    \dot{\Omega}(t)=\sum_l\dot{p}_l(t)\Omega_l\;,
\end{equation}
from which we can infer the radiated power (\ref{RotPower})
\begin{equation}\label{Quantized Power}
    P(t)=-\sum_l\hbar\omega_l\dot{p}_l(t)\;,
\end{equation}
by taking the negative time derivative of the energy
\begin{equation}
    E = \sum_l p_l(t)E_l=\sum_l p_l(t)\hbar\omega_l\;,
\end{equation}
where $E_l=\hbar\omega_l$. 

\textit{Spontaneous decay---}The spontaneous decay rate $\Gamma(t)$ can be computed from the transition probability $p_{n\neq i}(t)$ between the initial state $\ket{i}$ and final one $\ket{n}$ of the molecule--field composite system. As interaction Hamiltonian of this system we use $\hat{H}_{\mathrm{int}}=-\hat{\bm{d}}\cdot\hat{\bm{E}}(\bm{r}_{\text{M}})$ to be a constant perturbation turned on at time $t=0$. Here $\hat{\bm{d}}$ is the electric dipole operator and $\hat{\bm{E}}$ denotes the electric field operator evaluated at the molecule's position $\bm{r}_{\text{M}}$. Thus, at first order in time-perturbation theory the transition probability between the two states reads \cite{Sakurai}
\begin{align}
 p_{n\neq i}(t)=\frac{4|V_{ni}|^2}{(E_n-E_i)^2}\sin^2\bigg(\frac{\omega_{ni}t}{2}\bigg)\;,
\end{align}
where $V_{ni}$ denotes the transition matrix element, $E_n$ and $E_i$ are the respective energies, and $\omega_{ni}$ the corresponding frequency between the states.

We use as initial $\ket{i}=\ket{l}\ket{\left\{0\right\}}$ and final states $\ket{n}=\ket{l-1}\ket{1_\lambda(\bm{r},\omega)}$, and evaluate the matrix element using the Green tensor formalism to find 
\begin{equation}
    V_{ni} = \bra{1_\lambda(\bm{r},\omega)}\bra{l-1}-\hat{\bm{d}}\cdot\hat{\bm{E}}(\bm{r}_{\text{M}})\ket{l}\ket{\left\{0\right\}}\;,
\end{equation}
where we have expressed the states of the composite system via angular momentum states for the molecule and Fock states for the field, respectively. Writing the electric field in terms of bosonic creation and annihilation operators, this becomes \cite{DF1}
\begin{equation}
    V_{ni}=-\bm{d}_{l-1,l}\cdot\mathbb{G}_\lambda^{*}(\bm{r_\text{M}},\bm{r},\omega)\;,    
\end{equation}
where $\mathbb{G}_\lambda$ denotes the Green tensor. Summing over all possible states results in summing over the polarizations $\lambda$ and integrating over all possible positions $\bm{r}$ and frequencies $\omega$, we obtain the transition probabilities 
\begin{align}
    &\sum_{n\neq i}p_n(t)=\frac{4}{\hbar^2}\sum_\lambda\int_0^\infty\frac{\d\omega}{(\omega_{l,l-1}-\omega)^2}\sin^2\left[\frac{(\omega_{l,l-1}-\omega)t}{2}\right]\nonumber\\
    &\times\int\d^3r\bm{d}_{l-1,l}\cdot\mathbb{G}_\lambda^{*}(\bm{r_\text{M}},\bm{r},\omega)\cdot\mathbb{G}_\lambda(\bm{r_\text{M}},\bm{r},\omega)\cdot\bm{d}_{l,l-1}\;,
\end{align}
where we have used $\bm{d}_{l-1,l}^*=\bm{d}_{l,l-1}$, and expressed $\omega_{ni}$ and $E_n-E_i$ in terms of the transition frequency of the rotor $\omega_{l,l-1}$ and the photon's energy $\omega$. Using the integral relation for the Green tensor \cite{DF2}
\begin{align}
    &\sum_{\lambda=e,m} \int \mathrm{d}^3 s \mathbb{G}_\lambda (\bm{r},\bm{s},\omega)\cdot \mathbb{G}_\lambda^{*\top} (\bm{r}',\bm{s},\omega) &\nonumber\\
    &=\frac{\hbar \mu_0 \omega^2}{\pi} \mathrm{Im}\mathbb{G}(\bm{r},\bm{r}',\omega) \;,
\end{align}
the integral over $\bm{r}$ can be solved. We find
\begin{align}
    \sum_{n\neq i}p_n(t)&=\frac{4\mu_0}{\pi\hbar} \int_0^\infty\frac{\omega^2\d\omega}{(\omega_{l,l-1}-\omega)^2}\sin^2\left[\frac{(\omega_{l,l-1}-\omega)t}{2}\right]\nonumber\\
    &\times\mathrm{Tr}[\bm{d}_{l-1,l}\otimes\bm{d}_{l,l-1}\cdot\mathrm{Im}\,\mathbb{G}(\bm{r}_\mathrm{M},\bm{r}_\mathrm{M},\omega)]\;.
\end{align}

We perform rotational averaging on the transition probabilities, evaluating the dipole tensor as (See End Matter)
\begin{align}
     \bm{d}_{l, l-1}\otimes  \bm{d}_{l-1,l}&\equiv\frac{1}{2l+1}\sum_{m,m'}\bm{d}_{l, m,l-1,m'}\otimes  \bm{d}_{l-1,m,l,m'}\notag\\
     &=\frac{d^2l}{3(2l+1)}\mathbbold{1}\;,\label{dotimesd}
\end{align}
where we have summed the expression over the quantum numbers $m,\,m^\prime$ and introduced the factor $\frac{1}{2l+1}$ to ensure a normalized initial state obtaining an isotropic result. Finally, after substituting the free-space Green tensor \cite{DF2}
\begin{align}
    \text{Im}\,\mathbb{G}(\bm{r}_\text{M},\bm{r}_\text{M},\omega)=\frac{\omega}{6\pi c}\mathbbold{1}\;, \label{Free G}
\end{align}
the transition probabilities read
\begin{align}
    &\sum_{n\neq i}p_n(t) = \frac{2\mu_0 d^2}{3\pi^2\hbar c}\frac{l}{2l+1} \nonumber\\
    &\times \int_0^\infty\d\omega\frac{\omega^3}{(\omega_{l,l-1}-\omega)^2}\sin^2\left[\frac{(\omega_{l,l-1}-\omega)t}{2}\right]\;. \label{probabilities}
\end{align}

Let us first consider the short-time limit of Eq.~(\ref{probabilities}), which occurs when $t\ll 1/|\omega_{l,l-1}-\omega|$. Introducing an ultraviolet frequency cut-off $\omega_c=2m_\mathrm{e}c^2/\hbar\gg\omega_{l,l-1}$ with $m_e$ as the electron mass, the frequency integral leads to
\begin{equation}\label{Proba Short Time}
    \sum_{n\neq i}p_n(t) = \frac{\mu_0 d^2}{24\pi^2\hbar c}\frac{l}{2l+1}\omega_c^4 t^2\;,
\end{equation}
allowing us to deduce
\begin{equation}
    p_l(t) = 1 - \sum_{n\neq i}p_n(t) = 1 - \beta t^2\;,
\end{equation}
with
\begin{equation}
    \beta = \frac{d^2 \omega_c^4}{24\pi^2\hbar\ve_0 c^3}\frac{l}{2l+1}\;.
\end{equation}
On the other hand, in the Markovian limit $t\gg 1/|\omega_{l,l-1}-\omega|$, we can use the identity $\lim_{a\rightarrow\infty}\sin^2(ax)/(ax^2)=\pi\delta(x)$ to show that the probability satisfies \cite{DF2}
\begin{equation}\label{MarkovianProba}
    \dot{p}_l(t) = \Gamma_{l+1,l}p_{l+1}(t)-\Gamma_{l,l-1}p_l(t)\;,
\end{equation}
with transition rates given as
\begin{align}
    &\Gamma_{l,l-1}(\bm{r}_\text{M}) = \frac{2\mu_0}{\hbar}\omega_{l,l-1}^2 \nonumber\\
    & \times 
    \mathrm{Tr}\left[ \bm{d}_{l-1,l}\otimes\bm{d}_{l,l-1}\cdot\mathrm{Im}\,\mathbb{G}(\bm{r}_\mathrm{M},\bm{r}_\mathrm{M},\omega_{l,l-1}) \right]\;.\label{ratesT=0}
\end{align}

Performing again rotational averaging on the transition rates according to Eq.~(\ref{dotimesd}) and after substituting the free-space Green tensor (\ref{Free G}), the final expression for the transition rates reads
\begin{align}
    \Gamma_{l,l-1}&=\underbrace{\frac{\hbar^2d^2}{3\pi\ve_0 I^3c^3}}_{=:\Gamma_0}\frac{l^4}{2l+1}. \label{decayrates}
\end{align}
\textit{Discussion---}We now turn to analyze the consequences of the previous results for the dissipative torque $N$ at a quantum level. Using the rotational states (\ref{L2}) and substituting Eq.~(\ref{OmegaDot}) into Eq.~(\ref{LIOmega}), we obtain  
\begin{equation}
    N = I \langle\dot{\Omega}\rangle = I \sum_l\dot{p}_l(t)\Omega_l \;.
\end{equation}
In the short-time limit with initial state $\ket{l}$, Eq.~(\ref{Proba Short Time}) yields
\begin{equation}
    N= I \dot{p_l}\Omega_l = -2I\beta \Omega_l\, t = -2\frac{I d^2 \omega_c^4}{24\pi^2\hbar\ve_0 c^3}\frac{l}{2l+1}\Omega_l\,t\;.
\end{equation}
In this time regime the torque is linear in the angular velocity.
In contrast, for the Markovian limit the probability obeys Eq.~(\ref{MarkovianProba}). For times $t\ll1/\Gamma_{l,l-1}$, we may assume $p_l\simeq1$, $p_{\Bar{l}\neq l}\simeq0$, so the resulting expression for the torque reads
\begin{equation}
    N = -\frac{d^2}{3\pi\ve_0 c^3} \sum_l \frac{l^3}{(2l+1)(l+1)}\Omega_l^3\;,
\end{equation}
where Eqs.~(\ref{Quantized Omega}) and (\ref{decayrates}) have been employed. Therefore, in this regime the torque is cubic in the angular velocity.

To further understand the physical origin of this rotational quantum friction, we study its classical limit. First, we calculate the power carried by a photon emitted due to the rotational transition from the state $l$ to $l-1$ of the molecule
\begin{equation}\label{Power}
    P_l = \hbar\omega_{l,l-1}\Gamma_{l,l-1} = \frac{d^2 \hbar^4 l^5}{3\pi\ve_0c^3I^4 (2l+1)} \;,
\end{equation}
where Eq.~(\ref{decayrates}) and the corresponding eigenfrequencies have been used. Next, we take the classical limit $\hbar\rightarrow0$ and $l\rightarrow\infty$ simultaneously, by keeping $\hbar l$ constant, obtaining
\begin{equation}\label{Classical Power 1}
    P_{\mathrm{cl}} = \frac{1}{6\pi\ve_0c^3} \frac{d^2 \hbar^4 l^4}{I^4} 
    = \frac{d^2}{6\pi\ve_0c^3} \left(\frac{L}{I}\right)^4 \;.
\end{equation}
Here we identified $L=\hbar l$ as the angular momentum in the classical limit. The result agrees with our earlier classical dissipative power (\ref{Eq classical power}). This emphasizes that radiation reaction is the classical counterpart of spontaneous emission in the respective limit of large excitations.

Now we turn to the full quantum result of the radiated power (\ref{Quantized Power}). Through the quantized angular momentum $\Omega_l$ (\ref{Quantized Omega}), we recognize the factor $\hbar^4/I^4$ in Eq.~(\ref{Power}), indicating a $\Omega_l^4$-dependence of the power. Therefore, we may substitute Eq.~(\ref{Quantized Omega}) into Eq.~(\ref{Quantized Power}) to obtain
\begin{align}
    P_l& = \frac{d^2}{6\pi\ve_0c^3} \frac{l^3}{(l+1)^2(l+\frac{1}{2})}\,\Omega_l^4,
    \intertext{with the quantum correction}
    \frac{l^3}{(l+1)^2(l+\frac{1}{2})}&=
    \begin{cases}
        0 \quad\text{for}\quad l=0\;,\\
        \frac{1}{6}\quad \text{for} \quad l=1\;, \label{lcases}\\
        \frac{16}{45}\;\; \text{for} \quad l=2 \;.
    \end{cases}
\end{align}
Applying these results to the minimal example of a 3-level system, we find the following rotational velocity (\ref{RotationalVelocity}) to the fourth power:
\begin{align}
    \Omega^4(t)&=\sum_{l=0}^2p_l(t)\left(\frac{\hbar}{I}\sqrt{l(l+1)}\right)^4\notag\\
    &=\left(\frac{\hbar}{I}\right)^4\left\{\frac{192}{43}\e^{-\Gamma_0t/3}+\left(36-\frac{192}{43}\right)\e^{-16\Gamma_0t/5}\right\}.\label{Omega3Level}
\end{align}
The power is calculated using Eq.~(\ref{Quantized Power}):
\begin{align}
    P(t)&=-\sum_{l=0}^2\hbar\omega_l\dot{p}_l(t)\notag\\
    &=\frac{16}{43}\frac{\hbar^2\Gamma_0}{I}\e^{-\Gamma_0t/3}+\frac{1296}{215}\frac{\hbar^2\Gamma_0}{I}\e^{-16\Gamma_0t/5},\label{P3Level}
\end{align}
where the prefactor is given by $\frac{\hbar^2\Gamma_0}{I}=\frac{d^2}{3\pi\varepsilon_0c^3}\left(\frac{\hbar}{I}\right)^4$.
\begin{figure}[t]
    \centering
    \includegraphics[width=\linewidth]{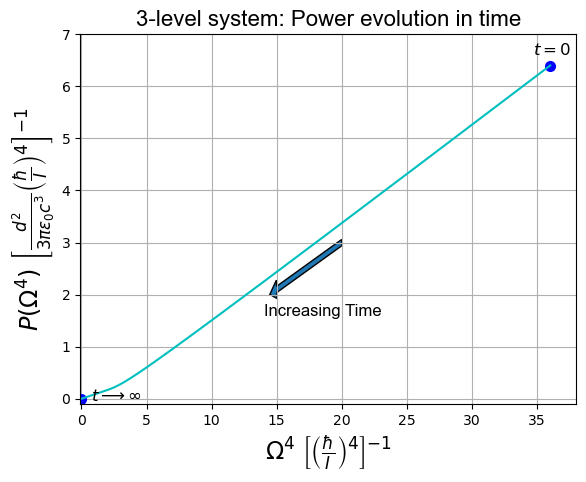}
    \caption{Evolution of the angular velocity $\Omega^4$ (\ref{Omega3Level}) and the radiated power $P$ (\ref{P3Level}) in time. The radiated power $P$ is plotted in units of $\dfrac{d^2}{3\pi\ve_0c^3}\left(\dfrac{\hbar}{I}\right)^4$, $\Omega^4$ in units of $\left(\dfrac{\hbar}{I}\right)^4$.}
    \label{Power evolution}
\end{figure}
By comparing $\Omega$ from Eq.~(\ref{Omega3Level}) and $P$ from Eq.~(\ref{P3Level}) in orders of $\hbar/I$, we may plot the resulting power as well as the quartic angular velocity $\Omega^4$ as function of time. For a 3-level system the result is shown in Fig.~\ref{Power evolution} for times $t\in[t=0,t\rightarrow\infty]$. Here, we find at $t=0$ a high initial angular velocity as well as a positive radiated power. Initially the power has a 
slope of $
16/45$ as seen from Eq.~(\ref{lcases}). As time increases, radiation is emitted, the slope decreases and the molecule is in a superposition of states with different $l$. In the long-time limit, the angular velocity and the power go to zero as the molecule reaches the equilibrium with the environment, which is reflected in a slope of $1/6$. This slope means that the transition probability from the state $l=1$ to the ground state $l=0$ dominates for times closer to $t_{\rightarrow\infty}$. 

Our analysis can be generalized to an environment at finite temperature $T$ where the transition rates are
\begin{align}
    &\Gamma_{l,l-1}(\bm{r}_\text{M})\notag\\
    \hspace{-1.5cm}&=\begin{cases}
        \frac{2\mu_0}{\hbar}\omega_{l,l-1}^2[n(\omega_{l,l-1},T)+1]\\
        \times\mathrm{Tr}\left[ \bm{d}_{l-1,l}\otimes\bm{d}_{l,l-1}\cdot\mathrm{Im}\,\mathbb{G}(\bm{r}_\mathrm{M},\bm{r}_\mathrm{M},\omega_{l,l-1}) \right],\hspace{.1cm} l\rightarrow l-1,\\
        \frac{2\mu_0}{\hbar}\omega_{l-1,l}^2n(\omega_{l-1,l},T)\\
        \times\mathrm{Tr}\left[ \bm{d}_{l-1,l}\otimes\bm{d}_{l,l-1}\cdot\mathrm{Im}\,\mathbb{G}(\bm{r}_\mathrm{M},\bm{r}_\mathrm{M},\omega_{l-1,l}) \right],\hspace{0.1cm} l-1\rightarrow l,
    \end{cases}\label{Temperature-dependent rates}
\end{align}
with the Bose--Einstein photon number being given by $\displaystyle n(\omega,T)=(\e^{\hbar\omega/k_\text{B}T}~-~1)^{-1}$. From Eq.~(\ref{Temperature-dependent rates}), one calculates a temperature-dependent friction coefficient $\alpha$ which can be shown to vanish in the long-time limit. We interpret this as the molecule's rotational energy matching the thermal energy of the vacuum, so in this long-term equilibrium no further friction is appreciable.

\textit{Conclusions and outlook---}In this Letter, we have established a framework for describing rotational quantum friction for a diatomic polar molecule undergoing a quantized motion. In the short-time limit, we found the friction to be linear in the angular velocity, while in the long-time Markov approximation we obtain a cubic velocity dependence. The resulting radiated power is in agreement with Larmor's formula in the classical limit. Furthermore, we studied temperature-dependent transition rates, which gave rise to a nonvanishing but constant rotation in the long-time limit,  meaning that the rotational and thermal energy reach equilibrium.

Our results provide a new perspective by understanding rotational quantum friction as a phenomenon that stems from real photons rather than virtual photons occurring in the Casimir-type and Casimir--Polder-type quantum friction \cite{Pendry1997,Pendry2010a,VolokitinRMP,Silverinha}.
A logical next step would be to investigate how to amplify this rotational friction via medium resonances, e.g.\ a dielectric plate, and sophisticated materials like topological insulators or chiral media \cite{OJF}. Such a scenario would provide a material-based frequency cut-off, so that the short-time dynamics could be more readily 
observed. Another interesting step would be to change the quantum emitter and analyze others like a nano-oscillator \cite{Aspelmeyer}.

\textit{Acknowledgements---}O.J.F. has been supported by the postdoctoral fellowship CONACYT-800966. The authors gratefully acknowledge funding by the Deutsche Forschungsgemeinschaft -- Project No. 328961117 -- SFB 1319 ELCH. 

\bibliography{Msc}


\begin{widetext}
\begin{center}
\large{\textbf{End Matter}}
\end{center}
\end{widetext}
\textit{Rotational Averaging---} When analyzing the rotational molecular dynamics in terms of decay rates, we do not resolve the orientation of the molecular rotation. Consequently, the molecular state is independent of the azimuthal quantum number $m$. The rotational eigenstates $\ket{lm}$, however, are anisotropic and depend on the choice of the quantization axis through $m$. To account for this, we perform an incoherent sum over all orientations for a given $l$, leading to the rotationally averaged density matrix
\begin{equation}\label{rho}
    \hat{\rho}=\frac{1}{2l+1}\sum_{m=-l}^l \ket{lm} \bra{lm}\;.
\end{equation}

We now evaluate the dyadic product $\bm{d}_{l,l-1}\otimes\bm{d}_{l-1,l}$ of the dipole transition moments. For a rigid rotor, the dipole operator is given by $\hat{\bm{d}}=d\bm{e}_r$ where $\bm{e}_r$ is the radial unit vector in spherical coordinates. The transition dipole moments between rotational eigenstates are defined as 
\begin{equation}
    \bm{d}_{l,m,l',m'} = \braket{ lm | \hat{\bm{d}} | l'm' }.
\end{equation}
Since the eigenstates $\ket{lm}$ are spherical harmonics, the matrix elements can be expressed as integrals over the solid angle. Writing $\bm{e}_r$ in terms of spherical harmonics and omitting their explicit angular dependence, we obtain
\begin{align}
    \bm{d}_{l,m,l',m'} &= d\sqrt{\frac{2\pi}{3}} \int \d\phi \d\theta \sin\theta \nonumber\\
    & \times Y^*_{lm}\left[
    \begin{array}{c}
        Y_{1-1}-Y_{11} \\
        \i\left(Y_{1-1}-Y_{11}\right) \\
        \sqrt{2}Y_{10}
    \end{array}
    \right] Y_{l'm'},
\end{align}
The problem thus reduces to evaluating integrals of the form $Y^*_{lm}Y^*_{1i}Y_{l'm'},$ with $i=1,0,-1$. Such integrals can be expressed in terms of Wigner 3$j$ symbols \cite{3j}. Using the identity $Y_{lm}=(-1)^m Y_{l,-m}$ and the selection rule $\Delta l=\pm1$, we restrict to downward transitions and set $l'=l-1$. Evaluating the integral involving $Y_{1,1}$ yields
\begin{align}
    &\int \d\phi \d\theta \sin\theta Y^*_{lm} Y_{11}Y_{l'm'} \nonumber\\
    & = (-1)^m \sqrt{\frac{3(2l+1)(2l-1)}{4\pi}} \nonumber\\
    &\times \left(
    \begin{array}{ccc}
        l & 1 & l-1  \\
        0 & 0 & 0
    \end{array}
    \right)\left(
    \begin{array}{ccc}
        l & 1 & l-1  \\
        -m & 1 & m'
    \end{array}
    \right)\;.
\end{align}
The magnetic quantum number selection rule $-m+i+m'=0$ implies $m=m'+1$ for $i=1$. Expressing the 3$j$ symbols in terms of Clebsch–Gordan coefficients and evaluating them explicitly \cite{CGC}, we obtain
\begin{align}
    &\int \d\phi \d\theta \sin\theta Y^*_{lm} Y_{11}Y_{l-1,m'} = (-1)^{2l-1+m-m'}\delta_{m,m'+1}\nonumber\\
    & \times\sqrt{\frac{3}{8\pi}}\sqrt{\frac{(l+m')(l+m'+1)}{(2l+1)(2l-1)}}\;.
\end{align}
Proceeding analogously, the remaining integrals read
\begin{align}
    &\int \d\phi \d\theta \sin\theta Y^*_{lm} Y_{1-1}Y_{l-1,m'} = (-1)^{2l-1+m-m'}\delta_{m,m'-1}\nonumber\\
    & \times\sqrt{\frac{3}{8\pi}}\sqrt{\frac{(l-m')(l-m'+1)}{(2l+1)(2l-1)}}\;,
\end{align}
\begin{align}    
    &\int \d\phi \d\theta \sin\theta Y^*_{lm} Y_{10}Y_{l-1,m'} = (-1)^{2l-2+m-m'}\delta_{m,m'}\nonumber\\
    & \times\sqrt{\frac{3}{4\pi}}\sqrt{\frac{(l+m')(l-m'+1)}{(2l+1)(2l-1)}}\;.
\end{align}

Collecting all contributions, the transition dipole moment takes the compact form
\begin{align}
    &\bm{d}_{l,m,l-1,m'} = \frac{d}{2} \frac{(-1)^{m-m'}}{\sqrt{(2l+1)(2l-1)}} \nonumber\\ 
    &\times\left[
    \begin{array}{c}
         -f_1(l,m) \delta_{m',m+1} + f_2(l,m) \delta_{m',m-1} \\
         -\i f_1(l,m) \delta_{m',m+1} -\i f_2(l,m) \delta_{m',m-1}  \\
         2 f_3(l,m) \delta_{m,m'}
    \end{array}
    \right]\;,
\end{align}
with $f_1(l,m)=\sqrt{(l-m')(l-m'+1)}$, $f_2(l,m)=\sqrt{(l+m')(l+m'+1)}$ and $f_3(l,m)=\sqrt{(l-m')(l+m')}$.

The other dipole $\bm{d}_{l-1,m,l,m'}$ follows by complex conjugation. Performing the incoherent sum over $m$ and $m'$ using the density matrix~(\ref{rho}), we find
\begin{align}
    &\bm{d}_{l,l-1}\otimes\bm{d}_{l-1,l} = \frac{1}{2l+1} \sum_{m,m'}\bm{d}_{l,m,l-1,m'}\otimes\bm{d}_{l-1,m,l,m'}\nonumber\\
    &=\frac{1}{2l+1}\frac{d^2}{4(2l+1)(2l-1)}\frac{4l(2l-1)(2l+1)}{3}\mathbbold{1}\nonumber\\
    &=\frac{d^2l}{3(2l+1)}\mathbbold{1}\;.
\end{align}

This result coincides with Eq.~(\ref{dotimesd}) in the main text and can equivalently be obtained by evaluating the expectation value  $\braket{\hat{\bm{d}}\otimes \hat{\bm{d}}}_l$.

\end{document}